\documentstyle[12pt]{article}
\begin{document}
\baselineskip 7mm
\title{ Proposal for an Experiment to Measure the Hausdorff Dimension
of Quantum Mechanical Trajectories }
\author{H. Kr{\"o}ger \\ [2mm]
{\small\sl D{\'e}partement de Physique, Universit{\'e} Laval,
Qu{\'e}bec, Qu{\'e}bec, G1K 7P4, Canada } \\
{\small\sl Email: hkroger@phy.ulaval.ca } }
\date{May 1995 \\
Univesit\'e Laval preprint: LAVAL-PHY-4/95}
\maketitle
\begin{flushleft}
{\bf Abstract}
\end{flushleft}
We make a proposal for a Gedanken experiment, based on the Aharonov-Bohm
effect,
how to measure in principle the zig-zagness of the trajectory of propagation
(abberation from its classical trajectory) of a massive particle in quantum
mechanics.
Experiment I is conceived to show that contributions from quantum paths
abberating from the classical trajectory are directly observable.
Experiment II is conceived to measure average length, scaling behavior and
critical exponent (Hausdorff dimension) of quantum mechanical paths.
\begin{flushleft}
PACS index: 03.65.Bz
\end{flushleft}
\setcounter{page}{0}

\newpage
\section{Background}
In order to demonstrate the distinction between classical mechanics
and quantum mechanics one often considers the ground state energy
of a system which has bound states. E.g., for the harmonic oscillator
classical and quantum mechanical ground state energy differ (in one space
dimension)
by
\begin{equation}
E_{qm} - E_{cl} = \frac{1}{2} \hbar \omega.
\end{equation}
The existance of such effect can be understood in terms of Heisenberg's
uncertainty principle.
But the distinction of classical and quantum mechanics does not only show up
in bound states but also in scattering states.
Let us consider the propagation of a massive particle.
Classical mechanics predicts that the particle follows smooth (differentiable)
trajectories. However, in quantum mechanics
according to Feynman and Hibbs \cite{kn:Feyn65} the paths are
non-differentiable, self-similar curves, i.e., zig-zag curves (see Fig.[1]).
Feynman and Hibbs have
noticed in 1965 the property of self-similarity, which plays an eminent r\^ole
in many areas of modern physics. Mandelbrot \cite{kn:Mand83} has introduced
the notion of fractal geometry and pointed out that self-similarity is a
characteristic feature of a fractal. Fractals are characterized by a fractal
dimension $d_{f}$ or a Hausdorff dimension $d_{H}$.
Abbot and Wise \cite{kn:Abbo81} have demonstrated by an analytic calculation
that quantum mechanical free motion yields paths which
are fractal curves with Hausdorff dimension $d_{H}=2$.
The corresponding classical system follows a trajectory with $d_{H}=1$
(the Hausdorff dimension $d_{H}$ coincides with the topological dimension
$d_{top}$ when the object is "not fractal" and $d_{top}=1$ for a curve).
Thus we can express the distinction between classical and quantum mechanics
by the Hausdorff dimension of propagation of a free massive particle,
\begin{equation}
d^{H}_{qm} - d^{H}_{cl} = 1.
\end{equation}
For the sake of later discussions, let us recall the basic ingredients of Abbot
and Wise's calculation. They consider the quantum mechanical motion of a free
particle. They assume measurements of position of the particle at consecutive
time intervals $\Delta t$. The uncertainty of position is $\Delta x$.
Let us call this a monitored path.
Taking a measurement means there is an interaction with the particle, hence it
is no longer "free". Without specifying what the interaction is, Abbot and Wise
consider that the measurement implies a minimal disturbance of momentum,
given by Heisenberg's uncertainty relation $\Delta p \ge \hbar/\Delta x$.
When $\Delta x$ goes to zero, this generates an erratic path. Their calculation
starts at $t=0$ from a localized $(\Delta x)$ wave packet and they compute
the length as expectation value of position at time $\Delta t$ and its scaling
as when $\Delta t \rightarrow 0$ and $\Delta x \rightarrow 0$. From this they
deduce $d_{H}=2$ for an average monitored path. This result has been
generalized by Campesino-Romeo et al. \cite{kn:Camp82} who find $d_{H}=2$ for a
monitored average path in the presence of a harmonic oscillator potential.

\bigskip

In contrast to that one can ask: What is $d_{H}$ for an unmonitored path? As
mentioned above, the geometric characteristics of quantum paths, like
zig-zagness, non-differentiability, and self-similarity have already been known
to Feynman and Hibbs by 1965.
It should be noted that they have also computed the essential pieces
which almost proves $d_{H}=2$ for an unmonitored average quantum mechanical
path. Their calculation includes the presences of any local potential.
Moreover, their calculation shows the close connection with Heisenberg's
uncertainty principle.
So let us recall here the basic steps of Feynman and Hibbs' calculation
\cite{kn:Feyn65a}.
They consider the Hamiltonian
\begin{equation}
H = \frac{- \hbar^{2} \Delta}{2 m} + V(x).
\end{equation}
where $V$ denotes a local potential.
The quantum mechanical transition element from a state
$\mid x_{in}, t=0>$ to a state
$\mid x_{fi}, t=T>$ is given by
\begin{equation}
< x_{fi}, t=T \mid x_{in}, t=0 > =
< x_{fi} \mid \exp[-\frac{i}{\hbar} H T ] \mid x_{in}> =
\int [d x(t)] \exp[ \frac{i}{\hbar} S[x(t)] ].
\end{equation}
The expression on the r.h.s. is the path integral \cite{kn:Feyn65},
i.e., the sum over
paths $x(t)$ which start at $x_{in}$ at $t=0$ and arrive at $x_{fi}$
at $t=T$. The paths are "weighted" by a phase factor $\exp[i S[x(t)]/\hbar ]$,
where $S$ is the classical action corresponding to the above Hamiltonian
for a given path,
\begin{equation}
S = \int_{0}^{T} dt \frac{m}{2} \dot{x}^{2} - V(x(t)).
\end{equation}
Analogously, the transition element of an operator $F[\hat{x}]$ is given by
\begin{equation}
< F[\hat{x}] > = < x_{fi}, t=T \mid F[\hat{x}] \mid x_{in}, t=0 > =
\int [d x(t)] \; F[x(t)] \; \exp[ \frac{i}{\hbar} S[x(t)] ].
\end{equation}
Now suppose time is divided in small slices $\delta$ ($x_{i}=x(t_{i})$ giving
the action
\begin{equation}
S = \sum_{i=1}^{i=N-1} \delta \left[ \frac{m}{2} \left(
\frac{x_{i+1}-x_{i}}{\delta} \right)^{2} - V(x_{i}) \right].
\end{equation}
Feynman and Hibbs obtain the general relation
\begin{equation}
< \frac{ \partial F }{ \partial x_{k} } > =
- \frac{i \delta }{\hbar} < F \frac{ \partial S}{\partial x_{k}} >.
\end{equation}
Putting $F=x_{k}$, this yields
\begin{equation}
<1> =  \frac{i}{\hbar}
< m x_{k} \left( \frac{ x_{k+1} - x_{k}}{\delta} -
\frac{x_{k} - x_{k-1} }{\delta} \right) + \delta x_{k} V'(x_{k}) >.
\end{equation}
In the limit $\delta \rightarrow 0$, the potential term becomes negligible
and one obtains
\begin{equation}
< m \frac{ x_{k+1} - x_{k}}{\delta} x_{k} > -
< x_{k} m \frac{x_{k} - x_{k-1} }{\delta} > = \frac{\hbar}{i}<1>.
\end{equation}
This means the transition element of a product of position and momentum
depends on the order in time of these two quantities. This leads to the usual
operator commutation law between position and momentum, which implies the
Heisenberg uncertainty
relation between position and momentum.
Now suppose one advances the second term on the l.h.s. of Eq.(10) by one time
slice $\delta$
\begin{equation}
< x_{k} m \frac{x_{k} - x_{k-1} }{\delta} > =
< x_{k+1} m \frac{x_{k+1} - x_{k} }{\delta} > + O(\delta)
\end{equation}
Then in the limit $\Delta t \equiv \delta \rightarrow 0$, Eqs.(10,11) imply
\begin{eqnarray}
&& < \left( x_{k+1} - x_{k} \right)^{2} > =
- \frac{\hbar \delta}{i m }<1>,
\nonumber \\
&& < ( \Delta x )^{2} > \; \propto \frac{\hbar }{ m } \Delta t.
\end{eqnarray}
This is Feynman and Hibbs' important result on the scaling relation
between a time increment $\Delta t $ and the corresponding average length
increment of a typical quantum path.
Now we make an assumption which we consider as plausible, but which we were not
able to prove
\begin{equation}
< \mid \Delta x \mid >^{2} = < (\Delta x)^{2} >.
\end{equation}

\bigskip

Now let us consider a finite time interval $T=N \Delta t$ and the length of the
path the particle has travelled between $x_{0}=x(t_{0})=x_{in}$ and
$x_{N}=x(t_{N})=x_{fi}$. Classically the length is given by
\begin{equation}
L_{class} = \sum_{k=0}^{N-1} \mid x_{k+1} - x_{k} \mid .
\end{equation}
According to Feynman and Hibbs, the corresponding quantity in quantum mechanics
is
the transition element given by Eq.(6), where $F[\hat{x}]$ is given by the
classical length,
\begin{eqnarray}
&  & < L > =
< x_{fi},t=T \left| \sum_{k=0}^{N-1} \mid x_{k+1}-x_{k} \mid \right| x_{in},t=0
> =
\nonumber \\
&  &
\int [d x(t)] \; \sum_{k=0}^{N-1} \mid x(t_{k+1})-x(t_{k}) \mid  \;
\exp[ \frac{i}{\hbar} S[x(t)] ].
\end{eqnarray}
One should note that this is not an expectation value in the usual sense.
Feynman and Hibbs refer to transition elements as "weighted averages".
The weighting function in quantum mechanics is a complex function.
Thus the transition element is in general complex.
Using Eqs.(12,13) one computes
\begin{equation}
<L> =  N <\mid \Delta x \mid> = \frac{T}{\Delta t} < \mid \Delta x \mid > \;
\propto \frac{T \hbar}{ m < \mid \Delta x \mid >}.
\end{equation}
Comparing this with the definition of the Hausdorff dimension (see Eqs.(17,18)
and putting $\epsilon = < \mid \Delta x \mid >$ yields Hausdorff dimension
$d_{H}=2$ for an unmonitored typical quantum mechanical path in the presence of
an arbitrary local potential.

\bigskip

Because the rigor of this result hinges upon the validity of the assumption in
Eq.(13).
it is interesting to check this result by a numerical calculation.
Kr\"oger et al. \cite{kn:Krog95} have computed the transition element
of the path length
$< L > = < \sum_{k=0}^{N-1} \mid x_{k+1}-x_{k} \mid >$
via numerical simulations of the path integral
on a time-lattice, however, using imaginary time (in order to be able to use
Monte Carlo methods).
The results for the Hausdorff dimension of unmonitored paths are compatible
with $d_{H}=2$ for free motion.

\bigskip

When we ask what is the Hausdorff dimension
for a quantum mechanical particle with interaction, we expect for local
potentials
via Feynman and Hibbs' calculation to obtain the value $d_{H}=2$.
The numerical simulations by Kr\"oger et al. \cite{kn:Krog95} in the presence
of
local potentials like harmonic oscillator or Coulomb potential
give results also compatible with $d_{H}=2$.
However, $d_{H} \neq 2$ has been found in the case of velocity dependent
interactions.
More precisely, for velocity dependent interactions
$U \sim U_{0} \mid \dot{x} \mid^{\alpha}$, the value $d_{H}=2$ has been found
for $\alpha \leq 2$, but
$d_{H} < 2$ for $\alpha > 2$.
Velocity dependent actions play a role in
condensed matter physics: The propagation in a solid medium
introduces higher order velocity terms via dispersion relations
\cite{kn:Made81}.
Also Brueckner's \cite{kn:Brue55} theory of nuclear matter saturation
introduces velocity dependent interactions.
The action relevant for this work, namely the interaction of a massive charged
particle with a vector potential, is also a velocity dependent action, being
linear in the velocity ($\alpha=1$). Thus one expects also $d_{H}=2$.

\bigskip

Let us summarize the present situation concerning the fractal dimension of an
average quantum path in the presence of a local potential:
A monitored path is a path where measurements of position are taken at some
discrete time intervals, i.e., the particle undergoes interaction. Possible
ways to do this are discussed in sect.3.
On the other hand an unmonitored path is undisturbed by interaction (in
non-relativistic quantum mechanics we neglect interaction with the vacuum,
particle creation, vacuum polarization etc.).
There is a rigorous proof of $d_{H}=2$ for monitored paths.
For unmonitored paths, there is strong indication that $d_{H}=2$ holds also.
However, neither a rigorous proof, nor a numerical simulation in real time has
been established so far. One might ask: What is the relationship between
monitored and unmonitored paths and why should $d_{H}$ coincide for both?
First of all, the operational definition of length employed is different.
For monitored paths, the authors of Ref.\cite{kn:Abbo81,kn:Camp82} have defined
length $\Delta l$ as the usual quantum mechanical expectation value of the
absolute value of an increment of position in the state of a wave function
having been evolved
for an increment of time $\Delta t$ from an original wave function being
characterized by localization uncertainty $\Delta x$.
In this work and Ref.\cite{kn:Krog95} we are interested in the length of
unmonitored paths, however, not in length
corresponding to an infinitesimal time interval, but the length corresponding
to a finite time interval, say $T$. This involves a number of intermediate
times.
The goal to seek information on the average of an observable at several times
leads to Feynman and Hibbs' transition element, which we have employed here.
Thus the length definition and average are different.
However, there is a common link, which may be considered as physical origin of
fractal paths: Heisenberg's uncertainty relation and behind this the
fundamental commutator relation between position and momentum.
For the case of the monitored paths, the process of localization leads via the
uncertainty principle to erratic paths. But also for the case of unmonitored
paths, Feynman and Hibbs' calculation shows that the scaling relation
between $\Delta x$ and $\Delta t$, Eq.(12), is directly related to the
commutator relation beteen position and momentum, which again is directly
related to the uncertainty principle.
But why then should the outcome of $d_{H}$ agree?
Because the result $d_{H}=2$ for unmonitored paths is not rigorously
established yet, one can only speculate on this hypothetical coincidence.
For unmonitored paths the scaling relation Eq.(12) is valid for a very large
class of interactions, namely all local potentials. Also the numerical
simulations in Ref.\cite{kn:Krog95} have given, within statistical errors,
$d_{H}=2$ for all local potentials investigated. In other words there is
indication (not a proof) that $d_{H}=2$ for unmonitored paths
in the presence of arbitrary local potentials. On the other hand, monitoring a
path means interaction by measurement. If one assumes that such interaction
is described by a local potential it seems plausible that $d_{H}$ coincides
for monitored and unmonitored paths.

\bigskip

If the zig-zagness of quantum paths is such a fundamental property of quantum
mechanics,
one might ask if it has been measured experimentally. To the author's knowledge
such an experiment has not been done yet.
Thus the central theme of this work are the following questions:
Can we observe experimentally
the zig-zagness of quantum mechanical trajectories?
Is such an experiment feasible in principle?
Can it be done in practice?
The motivation is twofold:
(a) Zig-zagness of paths is a fundamental property of quantum mechanics,
and thus it would be desirable
to have a direct experimental evidence. \\
(b) As we have mentioned above, there is indication
that velocity dependent interactions may change the Hausdorff dimension. Thus
an experiment measuring the Hausdorff dimension
would yield information on the interaction.

\bigskip

The zig-zagness of the free quantum mechanical motion
is principally not measurable,
because every measurement involves an interaction with the system.
Hence the system is no longer free.
Thus we can study at best the quantum mechanical motion
of an interacting particle. As our numerical simulations have shown, one would
stil expect a zig-zag motion.
For a certain class of potentials (local potentials) the fractal
dimension would be the same as for free motion.
What do we want to measure, in particular?
We want to measure the geometry of the average quantum mechanical
trajectory, in particular we want to measure the average length of the
trajectories, then the scaling of this length under variation
of an elementary length scale, and finally extract a critical exponent,
which is closely related to the fractal dimension (Hausdorff dimension) of the
trajectory.

\section{Reminder on fractal dimension}
A definition of the Hausdorff (fractal) dimension $d_{H}$ is given
by Mandelbrot \cite{kn:Mand83}.
He considers as example how to measure the length of the coastline of England.
One takes a yardstick, representing a straight line of a given length.
Let $\epsilon$ denote the ratio of the yardstick length to a fixed unit length.
Then one walks around the coastline, and measures the length of the coast with
the particular yardstick (starting a new step where the previous step leaves
off). The number of steps multiplied with the yardstick length
(characterized by $\epsilon$ ) gives a value $L(\epsilon)$
for the coastal length. Then one repeates the same procedure with a smaller
yardstick, say $\epsilon'$. Doing this for many values of $\epsilon$
yields a function $L$ versus $\epsilon$.
It has qualitatively the shape shown in Fig.[2].
One observes for a wide range of length scales $\epsilon$ that
the length of the British coast obeys a power law
\begin{equation}
L(\epsilon) = L_{0} \epsilon^{-\alpha}.
\end{equation}
This looks like the critical behavior of a macroscopic observable
at the critical point, thus $\alpha$ is called a critical exponent.
The Hausdorff dimension $d_{H}$ is defined by
\begin{equation}
\alpha = d_{H} - 1.
\end{equation}
So, one has an elementary length scale $\epsilon$, and one measures the length
of the curve $L(\epsilon)$.
Then $\epsilon$ is sent to zero.
One looks for a power law behavior (critical behavior) and determines the
critical exponent.
The Hausdorff dimension is directly related to the critical exponent.

\section{Measurement of position in quantum mechanics and elementary length}
In the calculation of the Hausdorff dimension for unmonitored paths in sect.1,
one has discretized time with an increment $\Delta t$.
The average increment of length $< |\Delta x| >$ and the average total length
$< L >$ have been determined dynamically by the system.
One has found a power law
$< L > \sim L_{0} < |\Delta x| >^{-\alpha}$ and the critical
exponent $\alpha$ has been extracted
in the limit $< |\Delta x| > \rightarrow 0$, which is equivalent to
$\Delta t \rightarrow 0$,
due to the scaling relation (12).
In an experiment, in order to measure the length $< L >$ and extract the
critical exponent
one would proceed differently: One would introduce an elementary length scale
$\Delta x_{exp}$ at the disposal of the experimentator.
Then one would measure the length $< L >$ by sequential measurements of
position as a function of $\Delta x_{exp}$, i.e., one would monitor the path.
Again one would expect a power law
$< L > \sim L'_{0} < \Delta x_{exp} >^{-\alpha'}$. Then one would approach the
limit
$\Delta x_{exp}  \rightarrow 0$. One would expect for the critical exponent
$\alpha' = \alpha$. A typical quantum mechanical path is sketched in Fig.[3].
Zig-zag motion occurs
in transversal as well as in longitudinal direction (not shown in Fig.).

\bigskip

We have to say what we mean by elementary length ("yardstick"). From the
experimental point of view it means the length scale of experimental resolution
when measuring position.
In the following we give several examples how this can be done:
(a) Sequence of absorbers (screens), each with a slit
(Fig.[4a]).
Here the elementary length $\Delta x$ is given by the distance between the
slits.
(b) Spark wire chamber (Fig.[4b]).
Here the elementary length $\Delta x$ is given by the distance between the
wires.
(c) Bubble chamber. Here the elementary length $\Delta x$ is given by the
spatial resolution of two different bubbles.
In the experimental proposal presented below  we will use neither of those, but
we will consider as elementary length the distance between neighbouring
solenoids carrying magnetic flux because the suggested experiment
is a generalized Aharonov-Bohm experiment.

\bigskip

How could one measure the length of a typical quantum path by monitoring the
position?
In principle, this can be done by a set-up of an electron source, a detector
and multiple screens with multiple holes (see Figs.[5a,b]).
In order to measure by which hole an electron has passed, Feynman and Hibbs
\cite{kn:Feyn65} have suggested two possible experiments:
(a) A light source is placed behind the screen (Fig.[5a]). One observes
scattering of light from the electron through one of the holes. From the
scattered light one can determine by which holes the electron has passed.
(b) As alternative, one can arrange such that the screen with holes
can move freely in vertical direction (Fig.[5b]). Before the electron passes,
the screen is at rest. When the electron passes, it scatters from the screen at
the hole.
The hole by which the electron passes can be determined by measuring if the
screen is recoiling up or downward, i.e., by measuring the momentum of the
screen.
One can imagine to place several such screens each carrying several holes.
Then one determines the holes which the electron has passed and clocks the time
for each passage of a hole, in order to determine if there has been
forward movement from the source to the detector.

\bigskip

The experimental length resolution $\Delta x$ is given by the distance between
neighbouring screens and the distance between holes on a screen.
We want to determine the length of the path. In order to extract a fractal
dimension we need to study $\Delta x \rightarrow 0$.
Then we run into the following problems:
The first one is a kind of technical problem:
When we put more screens between source and detector, then fewer electrons
will arrive at the screen. the counting rate goes down quite drastically.
This can be compensated by a longer beam-time, i.e., by emitting more
electrons.
There is, however, another more serious problem:
Consider alternative (a): In order to tell if an electron has passed by a
particular hole, the electron is scattered from light with wave length
$\lambda$. The (scattering) source of light of wave length $\lambda$
cannot be located in space with precision greater than order of $\lambda$.
Thus when we want to decrease the spatial resolution $\Delta x$, we need to
decrease the wave length $\lambda$ accordingly.
Light carries a momentum $2 \pi \hbar/\lambda$, which is (partially)
transferred to the electron. Thus the smaller is $\Delta x$, the larger becomes
the momentum transferred to the electron and the more the original quantum path
of the electron is altered. This is Heisenberg's uncertainty principle:
Any determination of the alternative (sequence of holes) taken by a process
capable of following more than one alternative destroys the interference
between alternatives. In other words: When we determine by which holes the
electron has passed, then the final interference pattern has no longer the
shape as shown in Fig.[6] (this has nothing to do with the fact that there are
several screens, it happens also for one screen with two holes).

\bigskip

For this reason, we are going to suggest below an experiment
without monitoring the path.
The experiment is different in the following sense: We do not place screens, so
there is no loss in the counting rate.
Secondly, our proposal is based on the Aharonov-Bohm effect,
which is classically a nul-effect, contrary to the above
set-up with screens which gives classically not a nul-effect.
Finally, in our proposal we determine the path via the topology
of the Aharonov-Bohm effect.

\section{Reminder on Aharonov-Bohm experiment }
The experiments we will suggest in the next sections are generalizations
of the Aharonov-Bohm effect. Thus it is worthwhile to recall
the notion of the Aharonov-Bohm effect and the experimental set-up, which has
confirmed this effect.
Aharonov and Bohm \cite{kn:Ahar59} have suggested in 1959 that there should be
an observable difference between classical and quantum mechanics
when a charged particle interacts with a magnetic field.
In particular, when the magnetic field (idealized) corresponds to an
infinitely long, infinitesimally thin flux tube (solenoid), then outside of the
flux tube the
magnetic field $\vec{B}=0$, but the vector potential is non-zero
($A_{\theta} \propto 1/r$ in spherical coordinates).
Due to the particular geometry of the magnetic field, which is independent of
the z-coordinate (the flux tube is assumed to be parallel to the z-axis),
we have an effective 2-D system in any plane perpendicular to the flux-tube.

\bigskip

Now consider a charged particle (charge $q$) passing by (scattering from) the
solenoid. Classically, the Lorentz force is zero.
But quantum mechanically, the electron wave function is affected by the
non-vanishing vector potential. Aharonov and Bohm \cite{kn:Ahar59}
have computed the cross section for scattering from an infinitesimally thin
flux tube with flux $\phi$. Choosing the gauge of the vector potential
such that the vector potential takes the form
\begin{equation}
A_{r}=0, \;\;\;  A_{\theta}= \phi /2\pi r,
\end{equation}
they find the cross section
(in the notation of Ref. \cite{kn:Wilc90}):
\begin{equation}
\frac{d \sigma}{d \theta} = \frac{1}{2 \pi k}
\frac{ \sin^{2}(\pi \alpha) }{ \sin^{2}(\theta /2) }, \;\;\;
\alpha = \frac{q \phi}{2 \pi \hbar c}.
\end{equation}
where $k$ is the wave number.

\bigskip

The Aharonov-Bohm effect has been confirmed by experiment \cite{kn:Wern60}.
The set-up corresponds to a two-slit interference experiment (see Fig.[6]).
A very thin solenoid is placed perpendicular to the plane
in the region between the two slits and the detector (region between the two
classical trajectories). Then the interpretation is as follows
\cite{kn:Feyn64}. One compares two cases:
(a) The solenoid is turned off (flux and vector potential are zero).
Then there is interference due to scattering from two sources. This is due to
a difference in the phase of the wave function
$\delta (\vec{B}=0)$, corresponding to the two trajectories.
(b) The solenoid is turned on (flux and vector potential are non-zero).
Then quantum mechanics says that the wave function
experiences a change of phase due to the presence of the vector potential
given by
\begin{equation}
\delta  =
\delta (\vec{B}=0) + \frac{ q }{ \hbar c} \int_{C} d \vec{l} \cdot \vec{A} ,
\end{equation}
where $C$ denotes the closed curve formed by the two trajectories,
$S$ is the area of the interior of this curve, $\vec{A}$ is the vector
potential, $\vec{B}$ is the magnetic field, $\phi$ is the magnetic flux
and $q$ is the charge of the particle.
Thus the phase change of the interference amplitude due to the presence of the
vector potential is basically given by
the flux through the region $S$,
\begin{equation}
\Delta \delta  = \delta - \delta(\vec{B}=0) =
\frac{ q }{ \hbar c} \int_{C} d \vec{l} \cdot \vec{A} =
\frac{ q }{ \hbar c} \int_{S} d \vec{s} \cdot \vec{B} =
\frac{ q \phi}{ \hbar c}.
\end{equation}
This change of phase $\Delta \delta$ shifts the maximum of the interference
pattern (see Fig.[6]). Again, classically there is no such effect,
because the magnetic field is practically zero
(exactly zero for an idealized solenoid) outside of the solenoid and hence
everywhere on the classical trajectory.
In summary, the Aharonov-Bohm effect shows that there is a difference
between quantum mechanics and classical mechanics, and moreover, the vector
potential is a real physical quantity.

\bigskip

In the Aharonov-Bohm experiment, as described above the quantum effects come
from the phase change of the wave function due to the presence of the vector
potential. In order to better understand the experiment suggested below, we
need to take a closer look into the quantum mechanics of the Aharonov-Bohm
effect.
It actually turns out that the above explanation in terms of phase change of
the wave function is valid only in the case when the distance $h$ between the
solenoid and the classical trajectories (see Fig.[6]) is large compared to the
de Broglie wave length $\lambda$. Note that the classical region is
given by $\Delta x >> \hbar/p = \lambda/2 \pi$, and the region of quantum
mechanics is given by $\Delta x << \hbar/p = \lambda/2 \pi$.
In general, the presence of the vector potential creates a phase change plus a
change of modulus of the wave function. This is so, in particular,
when one of the classical trajectories is close to the solenoid.
This is just the situation, which will play an important role in our
experimental proposal.

\bigskip

Schulman \cite{kn:Schu71} was the first to point out the connection between
quantum mechanics and topology in the Aharonov-Bohm effect. This holds in the
strict sense only in the
idealized situation of an infinitely thin and long flux tube.
Quantum paths can go by either to the left or to the right side of the
solenoid. Mathematically, this is equivalent to a plane with a hole
at the position of the solenoid. Quantum mechanical propagation proceeds
forward in time but forward and backward in both space dimensions (zig-zag
trajectories). Thus paths can occur, which wind around the solenoid (see
Fig.[7]). The classical Hamiltonian in the presence of the vector potential is
given by
\begin{equation}
H = \frac{1}{2m} \left( \vec{p} - \frac{q}{c} \vec{A} \right)^{2},
\end{equation}
and the action is given by
\begin{equation}
S = \int dt \; \frac{m}{2} \dot{\vec{x}}^{2} +
\frac{q}{c} \dot{\vec{x}} \cdot \vec{A}(\vec{x},t).
\end{equation}
Thus, when considering quantization by path integral, Eq.(6),
there occurs an Aharonov-Bohm phase factor due to the vector potential
present in the action.
This factor is $\exp[i \alpha (\theta_{fi}-\theta_{in} + 2\pi n_{w})]$,
(see Eq.(29) below), when the path winds $n_{w} = 0, \pm 1, \pm 2, \, \cdots $
times around the solenoid, and $\alpha = q\phi/2 \pi \hbar c$. This factor
depends only upon the initial and final
azimutal angle $\theta$ and the number of windings, but otherwise it is
independent of the path. In other words, paths can be classified by their
winding number, they fall into homotopy classes.

\bigskip

Now let us consider the Aharonov-Bohm propagator.
Fortunately, this expression can be computed analytically.
We follow the presentation by Wilczek \cite{kn:Wilc90}.
The idea is to start from the free propagator, decompose it into
classes corresponding to winding numbers $n_{w} = 0, \pm 1, \pm 2,
\cdots, \pm \infty$. Then one takes the free propagator in each winding class,
multiplies it by the Aharonov-Bohm phase factor and finally sums over all
windings.
The free propagator in $D=2$ dimensions is given by
(see Ref. \cite{kn:Schu81}),
\begin{eqnarray}
K^{free} &=& < \vec{x}_{fi} \mid \exp[ - i H T/\hbar ] \mid \vec{x}_{in} >
\nonumber \\
&=&
\frac{\mu}{2 \pi i \hbar T} \exp \left[ \frac{i \mu }{2 \hbar T}
(\vec{x}_{fi} - \vec{x}_{in})^{2} \right]
\nonumber \\
&=&
\frac{\mu}{2 \pi i \hbar T} \exp \left[ \frac{i \mu }{2 \hbar T}
(r'^{2} + r^{2} - 2 r'r \cos(\theta'-\theta)) \right].
\end{eqnarray}
In order to avoid confusion with the magnetic quantum number $m$,
the particle mass is denoted by $\mu$ in the following.
Now one allows $\theta'$ and $\theta$ correspond to different
winding number sectors. One defines $\Theta = \theta' -\theta +2 \pi n_{w}$.
The free propagator is periodic in $\theta$ between $-\pi$ and $\pi$.
Then one defines $\tilde{K}^{free}(\lambda)$
by Fourier transformation of $K^{free}(\theta)$,
\begin{eqnarray}
\tilde{K}^{free}(\lambda) &=& \int_{-\pi}^{+\pi} \frac{d \theta}{2 \pi}
\exp[ -i \lambda \theta ] K^{free}(\theta)
\nonumber \\
&=&
\int_{-\pi}^{+\pi} \frac{d \theta}{2 \pi}
\frac{\mu}{2 \pi i \hbar T} \exp \left[ \frac{i \mu }{2 \hbar T}
(r'^{2} + r^{2} - 2 r'r \cos(\theta)) \right]
\nonumber \\
&=&
\frac{\mu}{2 \pi i \hbar T} \exp \left[ \frac{i \mu }{2 \hbar T}
(r'^{2} + r^{2} ) \right] \; I_{| \lambda |}\left( \frac{\mu r r'}{i \hbar T}
\right),
\end{eqnarray}
where $I_{\nu}(z)$ is the modified Bessel function.
Thus the free propagator in the winding sector $n_{w}$
is given by
\begin{eqnarray}
&& K^{free}_{n_{w}}(\Theta)
\nonumber \\
&& = \int_{-\infty}^{+\infty} d \lambda \exp[ i \lambda \Theta ]
\tilde{K}^{free}(\lambda)
\nonumber \\
&& = \int_{-\infty}^{+\infty} d \lambda
\exp[ i \lambda (\theta' - \theta + 2 \pi n_{w}) ]
\frac{\mu}{2 \pi i \hbar T} \exp \left[ \frac{i \mu }{2 \hbar T}
(r'^{2} + r^{2} ) \right] \; I_{| \lambda |}\left( \frac{\mu r r'}{i \hbar T}
\right).
\end{eqnarray}
The total free propagator, being the sum over all windings is then
\begin{eqnarray}
&& K^{free}(r',\theta';r,\theta)
\nonumber \\
&& = \sum_{n_{w}=-\infty}^{+\infty}
\int_{-\infty}^{+\infty} d \lambda
\exp[ i \lambda (\theta' - \theta + 2 \pi n_{w}) ]
\frac{\mu}{2 \pi i \hbar T} \exp \left[ \frac{i \mu }{2 \hbar T}
(r'^{2} + r^{2} ) \right] \; I_{| \lambda |}\left( \frac{\mu r r'}{i \hbar T}
\right)
\nonumber \\
&& =
\sum_{m = -\infty}^{+\infty}
\exp[ i m (\theta' - \theta ) ]
\frac{\mu}{2 \pi i \hbar T} \exp \left[ \frac{i \mu }{2 \hbar T}
(r'^{2} + r^{2} ) \right] \; I_{| m |}\left( \frac{\mu r r'}{i \hbar T}
\right).
\end{eqnarray}
This is the free propagator, as given by Eq.(25), expressed in a more
complicated way. However, now it is easy to write down the Aharonov-Bohm
propagator. The Aharonov-Bohm propagator is the sum over all paths,
where each path is weighted with the phase $\exp[ i S/\hbar]$.
The paths can be decomposed into classes with a given winding number $n_{w}$,
where $n_{w}=0, \pm 1, \pm 2, \cdots$. Thus the Aharonov-Bohm propagator for a
a given winding number $n_{w}$ is just the free propagator for winding number
$n_{w}$ times the Aharonov-Bohm phase factor. For the vector potential
given by Eq.(19), this phase factor given by
\begin{equation}
\exp \left[ \frac{i}{\hbar} \int_{0}^{T} dt \frac{q}{c} \dot{\vec{x}} \cdot
\vec{A} \right]
= \exp \left[ \frac{iq}{\hbar c} \int_{x_{in}}^{x_{fi}} d\vec{x} \cdot \vec{A}
\right]
= \exp[ i \alpha (\theta' -\theta + 2\pi n_{w})].
\end{equation}
Thus one finds the Aharonov-Bohm propagator for winding number $n_{w}$
\begin{equation}
K^{AB}_{n_{w}}(\Theta) = \int_{-\infty}^{+\infty} d \lambda
\exp[ i (\lambda + \alpha) \Theta ]
\tilde{K}^{free}(\lambda).
\end{equation}
The total Aharonov-Bohm propagator is finally given by
\begin{eqnarray}
K^{AB}(r',\theta';r,\theta) &=& \sum_{n_{w}=-\infty}^{+\infty}
\int_{-\infty}^{+\infty} d \lambda
\exp[ i (\lambda + \alpha) (\theta' - \theta + 2 \pi n_{w}) ]
\tilde{K}^{free}(\lambda)
\nonumber \\
&=&
\sum_{m = -\infty}^{+\infty}
\exp[ i m (\theta' - \theta ) ]
\tilde{K}^{free}(\lambda-\alpha)
\nonumber \\
&=&
\sum_{m = -\infty}^{+\infty}
\exp[ i m (\theta' - \theta ) ]
\frac{\mu}{2 \pi i \hbar T} \exp \left[ \frac{i \mu }{2 \hbar T}
(r'^{2} + r^{2} ) \right] \; I_{| m -\alpha|}\left( \frac{\mu r r'}{i \hbar T}
\right).
\end{eqnarray}
The Aharonov-Bohn propagator, when letting $r', r \rightarrow \infty$
yields the Aharonov-Bohm differential cross section given by Eq.(20) (see
\cite{kn:Wilc90}. Inspection of the free propagator and the Aharonov-Bohm
propagator show the following structure: Apart from a common dimensionful
pre-factor $\mu/2 \pi i \hbar T$, they are both functions of dimensionless
arguments $\mu(r'^{2} + r^{2})/2 \hbar T$ and $\mu r' r /\hbar T$. The
Aharonov-Bohm propagator further depends on the dimensionless quantity
$\alpha = q \phi/2 \pi \hbar c$.
When we consider $r'=r$ large, then we have velocity $v=(r'+r)/T$, momentum
$p = \mu v$ and the de Broglie wave length
$\lambda = 2 \pi \hbar/p = \pi \hbar T/ \mu r$. Then the dimensionless argument
of the exponential function and the Bessel function becomes
$\pi r/\lambda$.

\bigskip

As mentioned above, the interpretation of the Aharonov-Bohm interference
experiment is based on the phase change of the wave function,
making the assumption that the solenoid is far away from the classical
trajectories. Because we have the exact expression for the Aharonov-Bohm
propagator at hand, we are able to test this assumption.
We introduce a semi-classical propagator, defined by
the product of the free propagator multiplied with the Aharonov-Bohm phase
factor computed along the classical trajectory (straight line between $x_{in}$
and $x_{fi}$, zero-winding)
\begin{equation}
K^{semi-class}(r',\theta';r,\theta) =
K^{free}(r',\theta';r,\theta)  \exp[ i \alpha (\theta' - \theta) ].
\end{equation}
Firstly, we have tested the free propagator expansion in terms of Bessel
functions.
We have compared the exact expression, Eq.(25), with the expansion in the
magnetic quantum number $m$, Eq.(28). We have imposed a cut-off $m_{max}$,
letting the sum run over $-m_{max} \leq m \leq m_{max}$. We have kept fixed the
values of parameters $\hbar=1$, $x_{in}$, $x_{fi}$, $L=2$, $T=10$, $\mu=1$.
$L$ is the length of a straight line between $x_{in}$ and $x_{fi}$, which is
the trajectory of classical propagation. We have chosen the origin
(of spherical coordinates) to be located at some distance $h$ from the
trajectory of classical propagation and equidistant from $x_{in}$ and $x_{fi}$.
We have varied $h$ from $0$ to $10$ and $m_{max}$ from $5$ to $15$. The
reference value, Eq.(25), is independent from $h$ and $m_{max}$. It has the
value $K^{free} = 0.00316129 - i \; 0.0155982$.
This set of parameters corresponds to the de Broglie wave length
$\lambda = 10 \pi$. If we identify the distance $h$ of the origin with the
resolution $\Delta x$, then we cross at $h=5$ from the quantum mechanical
region to the classical region.
Fig.[8a] shows the real part, Fig.[8b] corresponds to the imaginary part. One
observes rapid convergence in $m_{max}$.

\bigskip

We have tested the assumption on the asymptotic behavior of the Aharonov-Bohm
propagator by evaluating numerically
the Aharonov-Bohm propagator and the semi-classical propagator.
We have chosen the parameters as in Fig.[8],
now we have varied $h$ and $\alpha$.
The results are plotted in Figs.[9,10].
Fig.[9a] shows the real part of the semi-classical propagator, Eq.(32).
Fig.[9b] shows the real part of the Aharonov-Bohm propagator, Eq.(31). Here we
have chosen the cut-off $m_{max}=50$.
{}From the convergence test of the free propagator, we estimate that
$m_{max}=20$ should be sufficient to guarantee stability in the sixth
significant decimal digit when $h \leq 10$. The absolut value of the real part
of the difference between semi-classical and Aharonov-Bohm propagator is
displayed in Fig.[9c].
The corresponding results for the imaginary part are plotted in Fig.[10].
As in Fig.[8], this set of parameters corresponds to the de Broglie wave length
$\lambda = 10 \pi$ and the crossing of the quantum mechanical region to the
classical region occurs at $h=5$.
One observes that when the distance $h$ becomes large, the difference
between the Aharonov-Bohm propagator and the semi-classical propagator
tends to zero. Moreover, one observes that the difference between these
propagators is most pronounced for small distance $h$ ($h \rightarrow 0$).

\section{Gedanken experiment I to search for abberation from classical path
in quantum mechanical trajectories }
Now I want to suggest a Gedanken experiment in order to show
that not only there is a difference between classical and quantum mechanics,
but that quantum effects come from the zig-zagness of
quantum mechanical trajectories.
The set-up is based on the Aharonov-Bohm experiment.
But there is some modification.
Contrary to the original experiment, where the solenoid was placed
in the interior (center) of the region bounded by the
two classical trajectories, now I suggest to place the solenoid in the region
outside (see Fig.[11]).
Let $h$ denote the distance of the solenoid from the classical trajectory
$A-D$. Then we let the solenoid approach the classical trajectory
($h \rightarrow 0$).
Again we consider the case (a) when the solenoid is turned off and (b) when the
solenoid is turned on. The Gedanken experiment is to measure the
change of the interference pattern between cases (a) and (b).

\bigskip

Let us discuss what results we expect to find in this scenario, based on our
knowledge of the Aharonov-Bohm effect and the numerical calculation
of the propagator.
Firstly, let $h$ be large compared to the de Broglie wave length $\lambda$.
When $h$ is large, the Aharonov-Bohm propagator and the semi-classical
propagator coincide. Then like in the Aharonov-Bohm experiment, the wave
function of an electron experiences a change of phase, given by Eq.(21).
However, the phase change of the interference amplitude, given by Eq.(22),
now has the outcome $\Delta \delta =0$, because there is no magnetic flux
through the region $S$ (between the classical trajectories). Thus switching
off and on the magnetic field will not produce a change in the interference
pattern.

\bigskip

Now suppose the solenoid approaches the classical trajectory $A-D$, but we
assume that its distance from the other classical
trajectory $B-D$ is still large.
Then the propagator corresponding to the trajectory $A-D$ is given by the
Aharonov-Bohm propagator, while the propagator corresponding to the
trajectory $B-D$ is given by the semi-classical propagator.
Our previous numerical results show a marked difference between the
Aharonov-Bohm propagator and the semi-classical propagator (for propagation
between the same points).
There is not only a difference in phase, but also in modulus.
Thus we expect for the interference experiment, when $h$ becomes small and when
the magnetic field is switched on and off, that a change of phase as well as of
modulus will show up in the interference amplitude. Its interpretation is as
follows:
Quantum mechanics, expressed in the language of path integrals tells us that
there are contributions from all possible paths. There paths close to the
classical path (see Fig.[12]), which give dominant contributions.
There are paths far away from the classical path which give very small
contributions. There are also paths crossing the area of the solenoid
(in the case of finite extension).
As long as all paths of propagation between slit $A$ and detector $D$
pass by the same side of the solenoid (see Figs.[11,12a]), there will be zero
change in phase and modulus in the interference amplitude.
But when paths occur, which pass on both sides of the solenoid (Fig.[12b]), or
eventually wind around the solenoid as quantum mechanics predicts,
then interference is described by the Aharonov-Bohm propagator for
route $A-D$ and by the semi-classical propagator for the route $B-D$,
producing a change of the interference pattern.
In other words, any difference between the interference patterns for magnetic
fields $\vec{B}=0$
and $\vec{B}\neq 0$ indicates that there must be contributions from paths which
have deviated from the classical path $A-D$ by at least a distance $h$.
Thus the proposal of experiment I is:
(1) Measure the interference pattern when $\vec{B}=0$.
(2) Measure the interference pattern when $\vec{B} \neq 0$,
as a function of $h$.
Any difference between (1) and (2) signals that quantum paths have fluctuated
at least by a distance $h$ from the classical trajectory.

\bigskip

In Figs.[9,10] we have presented results in dimensionless units.
Now we want to take a look at the behavior of the propagator
for a more physical choice of parameters.
We take $\hbar=c=1$, thus $\hbar c = 197 MeV \; fm =1$, and express
every quantity in (powers of) $fm$. The electron mass is $\mu_{e} =
0.511 MeV/c^{2} = 0.259 \; 10^{-2} fm^{-1}$. We choose
$L = 2 \; cm = 2 \; 10^{13} fm$,
$T = 1 \; s = 3 \; 10^{23} fm$. The diameter of an elementary solenoid
used in an Aharonov-Bohm experiment is in the order of
$d_{sol} \sim 5 \; \mu m =5 \; 10^{9} fm$. We have varied
$h$ from $h = d_{sol} = 5 \; 10^{9} fm$ to
$h = 100 \; d_{sol} = 5 \; 10^{11} fm$.
This set of parameters corresponds to a de Broglie wave length
$\lambda = 3.63 \; 10^{13} fm$. If we again identify $h = \Delta x$,
then we have
$2 \pi \Delta x/ \lambda = 0.86 \; 10^{-3}$ for $h = d_{sol}$ and
$2 \pi \Delta x/ \lambda = 0.86 \; 10^{-1}$ for $h = 100 \; d_{sol}$. Thus we
are in the quantum mechanical region.
Experimentally, the magnetic flux can be varied continuously in a certain
range \cite{kn:Tono83}.
Dirac's quantization condition of flux from a Dirac-string attached to a
magnetic monopole is given by \cite{kn:Chen84}
$q \phi / 2 \pi \hbar c = \alpha = n, \;\; n=0,1,2, \cdots$
Thus it is reasonable to vary $\alpha$ in the order of unity.
We have varied $\alpha$ from $0$ to $3$.
We have plotted the propagators after dividing by the common factor
$\mu / 2 \pi i \hbar T$. Fig.[13a] displays the real part of the semi-classical
propagator, Fig[13b] shows the real part of the Aharonov-Bohm propagator
and in Fig.[13c] we have plotted the absolut value of the real part of their
difference.
Fig.[13c] shows that when the solenoid has a distance from the classical
trajectory in the order of several $d_{sol}$, there is a marked difference
in particular for $\alpha \approx 1/2,3/2,\cdots$, i.e. half integer.
This quantum mechanical effect should be observable experimentally.

\bigskip

Summarizing this section, a change in the interference pattern can only occur
due to contributions of paths, which abberate from the classical path,
in such a way that the solenoid is in the interior of the region
bounded by the two paths (coming from the two slits).
Thus the observation of a change of the interference pattern
in this Gedanken experiment shows directly the necessity to
take into account contributions from paths abberating from the classical path.
But it does not necessarily show that the paths are zig-zag.
The abberation could be a smooth one like a $sin$ curve.
In order to get more information on the geometry of those paths,
in particular on average length, scaling behavior, critical exponent,
i.e., Hausdorff dimension, we suggest another Gedanken experiment.

\section{Gedanken experiment II to measure the fractal dimension of quantum
mechanical trajectories}
\subsection{Set-up}
This set-up is a generalisation of the set-up of Gedanken experiment I.
As discussed above, one needs an elementary length scale $\Delta x$ and one
has to measure the length of the trajectory in terms of this
elementary length scale.
I suggest to take as elementary length scale
the distance between two neighbouring solenoids in an array of solenoids
(see Fig.[14]). The array of solenoids is placed such that the classical
trajectory coming from slit A, passes through this array,
while the classical trajectory coming from the slit B,
does not pass through this array. For example, the array could be placed in the
lower half plane bounded by a line of points which are
equidistant from the two slits.

\bigskip

As pointed out above, analytical and numerical calculations
have given Hausdorff dimension $d_{H}=2$
for free quantum mechanical motion. This is valid for motion in $D=1,2,3$
space dimensions. In this sense the trajectory shown in Fig.[3] is
not quite correct: The motion is zig-zag in longitudinal as well as in
transversal direction, while in Fig.[3] zig-zagness is shown only in
transversal direction. This experiment will be sensitive only to zig-zagness in
the plane perpendicular to the flux tubes.
The reason is as follows:
The massive charged particle interacts with the vector potential
corresponding to the magnetic field of the solenoids.
The solenoids represent (idealized) infinitely long and thin flux tubes.
The system is invariant with respect to translations parallel
to the flux tubes. Any zig-zagness of motion in this direction
does not show up in the phase factor
$\exp[ iq/\hbar c \int_{C} d \vec{x} \cdot \vec{A} ]$
and hence does not show up in the interference pattern.

\bigskip

{}From Gedanken experiment I we have learnt that observable quantum effects
occur when quantum paths deviate from the classical path such that
quantum paths go by both sides of the solenoid. As a consequence,
this experiment is sensitive only to zig-zagness on a
length scale larger than the minimal distance between two solenoids, i.e.,
$\Delta x$. This is due to the topological character of our experiment.
Thus we ask: What are the topologically different (homotopy) classes of paths,
corresponding to the given geometry of selonoids assuming we have $N_{S}$
solenoids
positioned in a regular array with next neighbour distance $\Delta x$?
The topological class of a path depends on the starting point $x_{in}$, the
endpoint $x_{fi}$ and the way it winds around the individual solenoids.
One should note that this does not depend on the sequential order
of winding around individual solenoids. Equivalent paths with
the same winding, but different sequential order are shown in Fig.[15].
Mathematically, this is characterized by the phase factor
$\exp[ iq/\hbar c \int_{C} d \vec{x} \cdot \vec{A} ]$, corresponding to
the path $C$.
If, e.g., the path $C$ is closed and winds around solenoids $1, \cdots, N_{S}$,
respectively, with winding numbers $n_{1}, \cdots, n_{N_{S}}$, this phase
factor yields
\begin{equation}
\exp \left[ \frac{iq}{\hbar c} [n_{1} \phi_{1} + \cdots n_{N_{S}} \phi_{N_{S}}
] \right].
\end{equation}
{}From the knowledge of this phase factor one can extract the winding numbers
of
solenoids in a unique way, if
the flux values $\phi_{1}$ to $\phi_{k}$ are incommensurable, i.e.,
their ratios are not rational numbers. In practice, however, quantum effects in
the amplitude due to high winding numbers become very small and eventually
for $n_{w}$ larger than some cut-off $n_{cut-off}$ they are no longer
detectable in an experiment. Thus we can allow the ratio of fluxes
\begin{equation}
\frac{\phi_{i}}{\phi_{j}} = \frac{n_{i}}{n_{j}}, \; n_{i}, n_{j} > n_{cut-off}.
\end{equation}

\bigskip

The experiment consists of measuring the interference pattern,
once when all solenoids are turned off and once when all solenoids are turned
on. Any change in the interference pattern is due to a
change of wave function which traverses the array of solenoids.
How is the change of the phase and modulus of the wave function related to the
flux of the solenoids? This is answered by
the rules of quantum mechanics expressed in terms of the path integral:
The wave function is given by the path integral
\begin{equation}
\psi(\vec{x},t) =
\left. \int [dx] \; \exp[ i/\hbar S[\vec{x}]]
\right|_{\vec{x},t;\vec{x_{0}},t_{0}},
\end{equation}
representing all paths between $\vec{x}_{0},t_{0}$ and $\vec{x},t$.
For the purpose of the interference experiment one would choose
$\vec{x}_{0},t_{0}$ corresponding to the source and $\vec{x},t$
corresponding to the detector. All paths go through either one of the two
slits.
The integral as a sum over paths.
By splitting the action into a kinetic and a magnetic part the wave function
can be expressed as
\begin{equation}
\psi(\vec{x},t) = \left.
\sum_{C} \; \exp[ i/\hbar S_{free}[C]
\; \exp[ \frac{iq}{\hbar c} \int_{C} d\vec{x} \cdot \vec{A}(\vec{x},t) ]
\right|_{\vec{x},t;\vec{x_{0}},t_{0}}.
\end{equation}
The wave function is a superposition of phase factors.
Each phase factor has been split into a term representing the weight of the
free action and into a term representing the phase of a line integral
of the vector potential. If we would retain in Eq.(36)
only one path $C_{1}$ passing through slit 1
and one path $C_{2}$ passing through slit 2 (which is of course unphysical),
then the interference pattern, given by the absolute amplitude squared would be
determined by the difference of the phases corresponding to $C_{1}$ and $C_{2}$
(see Ref.\cite{kn:Feyn64}). But the difference of two line integrals, both
going from $\vec{x}_{0},t_{0}$ to $\vec{x},t$
corresponds to a line integral along a closed curve $C$ (following $C_{1}$
and returning by $C_{2}$).
Then the second phase factor in Eq.(36) would describe the total flux going
through the area $S$ interior to $C$, as given in Eqs.(21,22).

\subsection{Reconstruction of paths: case of two paths}
The Gedanken experiment measures the interference pattern and hence the change
of the phase and modulus of the wave function. The question is how to extract
from this information about the geometry of paths, and in particular about
the length of the average path.
We want to treat this problem in two steps: Firstly, suppose there is only one
path corresponding to each slit. The shift of the phase $\Delta \delta$
of the wave function is then given by the sum of fluxes of the solenoids in the
interior area $S$.
In Fig.[14], those are the fluxes of solenoids 1 to 6, 10, 11, 13 and 14.
Suppose we know the shift $\Delta \delta$ and hence the total flux.
How can we tell from that which particular solenoids have contributed
to the flux? This information is necessary in order to trace the trajectory.
It can be answered in the following way:
We must assign to each solenoid a particular flux,
such that knowing the total flux allows to reconstruct which individual
fluxes have contributed. This is certainly not possible, when all
individual fluxes have the same value.
This problem is mathematically equivalent to
the following problem in number theory:
Given is a finite set of real numbers $r_{i}, i=1,\cdots,N$, and a real number
$R$. Suppose we know that the equation
\begin{equation}
n_{1} r_{1} + \cdots n_{N} r_{N} = R
\end{equation}
possesses a solution for a set of integer numbers $n_{1}, \cdots, n_{N}$.
We want to know: Under what conditions is this solution unique?
A possible answer is the following:
One chooses $r_{i}, i=1, \cdots, N$ as ratio of large integer numbers, which do
not possess a common integer divisor. One
imposes a cut-off on the integer numbers $n_{i} < n_{cut-off}$, where
$n_{cut-off}$ is small compared to the integers occuring in the numerators and
denominators of $r_{i}$. Then the solution is unique.
Let us give an example: $r_{1}=97/99$, $r_{2}=101/111$, $R=8463/10989$.
Then
\begin{equation}
\frac{8463}{10989} = n_{1} \frac{97}{99} + n_{2} \frac{101}{111}
\end{equation}
possesses the solution $n_{1}=-2$, $n_{2}=3$, but no other integer solution for
say $-3 < n_{i} < 3$.

\bigskip

Here we have made the following association between
the integer numbers and our experiment:
\begin{eqnarray}
R & \leftrightarrow & \mbox{ total flux }
\nonumber \\
r_{j} = & \leftrightarrow & \mbox{ flux of individual solenoid } j
\nonumber \\
n_{j} = & \leftrightarrow &
\mbox{ winding number of path around solenoid j }
\end{eqnarray}
Suppose we have tuned the magnetic field of the individual solenoid $j$,
such that the corresponding individual flux takes on the
value $r_{j}$. From the experiment we know the total flux $R$.
Thus we can determine in a unique way the winding numbers $n_{j}$.

\subsection{Reconstruction of paths: case of many paths}
The above  procedure allows to reconstruct the path, if there is a single path.
In quantum mechanics there are infinitely many paths which contribute
to the amplitude.
How can we reconstruct the average path from the knowledge of the
interference pattern in the general case?
The idea is to take advantage of the connection between quantum mechanics and
topology. Recall the discussion of the Aharonov-Bohm effect:
The quantum mechanical free propagator can be written as a decomposition of the
free propagator into winding number sectors and then summing over all winding
numbers (Eq.28).
The Aharonov-Bohm propagator
has a similar decomposition. However, now the contribution from a given
winding sector is the product of the free propagator
in this winding sector times the Aharonov-Bohm phase factor
$\exp[i \alpha (\theta'-\theta + 2\pi n_{w})]$ (Eq.31).
One aspect of this is important in the following:
If one varies the magnetic flux in the solenoid $\phi \rightarrow \phi'$ and
hence $\alpha \rightarrow \alpha'$, this changes of course the total
Aharonov-Bohm propagator. It also changes the Aharonov-Bohm
phase factor in each winding sector. But it {\it does not change}
the free propagator in each winding sector.

\bigskip

All this carries over to the generalized Aharonov-Bohm setting with
the array of $N_{S}$ solenoids. Again the full propagator decomposes into
homotopy classes.
The contribution from each homotopy class is the free propagator in this
homotopy class times a generalized Aharonov-Bohm phase factor,
\begin{equation}
\exp \left[ \frac{i q}{2 \pi \hbar c}
[ (\theta' -\theta) \phi_{tot}
+ 2\pi [ n_{1} \phi_{1} + \cdots n_{N_{S}} \phi_{N_{S}} ] \right],
\;\;\; \phi_{tot} = \phi_{1} + \cdots \phi_{N_{S}}.
\end{equation}
The important aspect is again: Changing the fluxes $\phi_{i}$
changes the full propagator, but does not change
the free propagator in each homotopy class.

\bigskip

Thus experimentally, we have a handle
to determine the free propagator corresponding to a given homotopy class.
We introduce a cut-off in the winding numbers $n_{i} < n_{cut-off}$.
What value would one attribute to $n_{cut-off}$?
In a Gedanken experiment one considers an idealized situation neglecting
experimental errors. In such a situation one would consider $n_{cut-off}$ as a
parameter which should be increased until the Hausdorff dimension of the
quantum path (sect.6) converges. In a more realistic situation one faces
experimental errors or more precisely thresholds (limits) of experimental
resolution. One expects that
winding numbers beyond the cut-off
give contributions to the amplitude which are in the order of experimental
errors and hence can not be detected.
The relation between a given experimental threshold and the corresponding
value of $n_{cut-off}$ could be estimated via a numerical simulation of the
path integral.

\bigskip

This cut-off makes the number of homotopy classes finite. Let us enumerate the
homotopy classes by $n_{h}=1,2, \cdots, N_{H}$.
The experimantator chooses a set of fluxes of the solenoids:
$\phi^{(1)}_{i}, i=1,\cdots,N_{S}$. Then he measures the corresponding
interference pattern, say $I^{(1)}$.
Then the experimentator chooses another set of fluxes of the solenoids,
$\phi^{(2)}_{i}, i=1,\cdots,N_{S}$, and measures again the interference
pattern,
$I^{(2)}$. This is repeated for $N_{F}$ differents sets of fluxes.
The gathered information is then sufficient to determine the
free propagators in the homotopy classes $n_{h}=1, \cdots, N_{H}$.
The wave function for emission from the source at $\vec{x}_{0}, t_{0}$
is given by
\begin{eqnarray}
\psi(\vec{x},t)
&=& \left.
\sum_{C} \; \exp[ i/\hbar S_{free}[C]
\; \exp[ \frac{iq}{\hbar c} \int_{C} d\vec{x} \cdot \vec{A}(\vec{x},t) ]
\right|_{\vec{x},t;\vec{x_{0}},t_{0}}
\nonumber \\
&=& \sum_{n_{h}}
K^{free}_{n_{h}}
\exp \left[ \frac{i q}{2 \pi \hbar c}
[ (\theta' -\theta) \phi_{tot}
+ 2\pi [ n_{1} \phi_{1} + \cdots n_{N_{S}} \phi_{N_{S}} ] \right].
\end{eqnarray}
The interference pattern is given by the squared modulus of the wave function
\begin{equation}
I = \mid \psi(\vec{x},t) \mid^{2}.
\end{equation}
Thus considering $N_{F}$ different sets of fluxes, one has
\begin{equation}
I^{(f)} = \left|
\sum_{n_{h}}
K^{free}_{n_{h}}
\exp \left[ \frac{i q}{2 \pi \hbar c}
[ (\theta' -\theta) \phi^{(f)}_{tot}
+ 2\pi ( n_{1} \phi^{(f)}_{1} + \cdots n_{N_{S}} \phi^{(f)}_{N_{S}} ) ] \right]
\right|^{2}, \;\;\; f=1,\cdots,N_{F}
\end{equation}
Because a given set of fluxes and a given homotopy class $n_{h}$
determines the generalized Aharonov-Bohm phase factor,
and the free propagator in each homotopy class $K^{free}_{n_{h}}$
is independent from the fluxes,
this equation allows to determine the unknown coefficients
$K^{free}_{n_{h}}, \;\;\; n_{h}=1,\cdots,N_{H}$.
Because $K^{free}_{n_{h}}$ are complex numbers, and the fluxes $\phi$ and
vector potentials $\vec{A}$ are real, we need at least twice as many sets of
fluxes as
the number $N_{H}$ of homotopy classes considered, $ N_{F} > 2 N_{H} $.

\subsection{Length of paths}
Suppose we know by now from experiment the values of $K^{free}_{n_{h}}$
corresponding to given a homotopy class $n_{h}$.
What length of path would we assign to it?
Consider the array of solenoids with distance
$\Delta x$ between neighbouring solenoids.
Homotopy classes are distinguished
by the phase factor of the vector potential
$\exp[ i q/\hbar c \int_{C} d\vec{x} \cdot \vec{A} ]$.
We start by grouping paths occuring in the path integral, Eq.(36),
into classes: Two paths are in the same class, if they can be made identical by
stretching and deforming without crossing a solenoid. This reflects
the fact that we can not resolve any structure on a scale smaller than
the resolution $\Delta x$. Moreover, the phase factor of the vector potential
does not distinguish the paths, like e.g., in Fig.[15a,b]. These two paths
belong to the same class. These two paths can be transformed into each other by
deformation of paths and applying the following rule: When two oriented paths
cross each other, they can be cut at the vertex such that each each piece has
an incoming and an outgoing line relative to the vertex.
For each class we define a representative path, which passes
in the middle between two neighbouring solenoids. Then one draws a straight
line from the middle of one of those pairs to the middle of the next of those
pairs (see Fig.[16]). Each of those representative paths is defined by a given
sequence of pairs of solenoids.
What representative path do we associate, e.g., to those topologically
equivalent paths of Fig.[15a,b]? Because the length of the curve
is expected to go to infinity, when $\Delta x \rightarrow 0$,
we are experimentally on the conservative side, if we choose the shortest path
from those of Fig.[15a,b], that is, the path which goes forward in longitudinal
direction, except for loops around individual solenoids (this is shorter than
doing one loop around two solenoids).
Thus the rule for representative paths is:
It starts at $x_{in}$ and arrives at $x_{fi}$. It goes by pieces of straight
lines always passing in the middle of a pair of solenoids.
The representative path is determined by the winding numbers
corresponding to the solenoids of the array. Among several paths
compatible with the same winding numbers, the shortest path is taken as
representative path. That means, each loop goes around at most one solenoid
(possibly several times).
An example of such a path is shown in Fig.[16]. In this sense, we associate to
each homotopy class a representative path. Let us denote those representative
paths by $C_{r}$.
Then we define the classical length of the trajectory as follows:
\begin{equation}
L_{C_{r}}(\Delta x) = \mbox{ sum of length of pieces of straight line }.
\label{eq:classdeflength}
\end{equation}
Finally, we want to obtain a quantum mechanical expression for
the length of paths. The general expression is a weighted average given by
the transition element
\begin{equation}
<L> = \frac{ \sum_{C} L_{C} \exp[i S[C] /\hbar ] }
           { \sum_{C} \exp[i S[C]/\hbar ] },
\label{eq:gendeflength}
\end{equation}
where the sum goes over all possible paths $C$ which go from the source at
$t_{in}$ to the detector at $t_{fi}$.
This is meant to be the same length definition as given in Eq.(15), but taking
$N \rightarrow \infty$ and introducing a normalizing factor.
In the discussion of experiment II we have introduced an elementary length
scale $\Delta x$.
In order to obtain an operational length definition for the experiment,
we replace the infinite sum over all paths in Eq.(~\ref{eq:gendeflength})
by a finite sum over the representative paths and the weight factor
$\exp[ i S/\hbar ]$ is given by the action in the homotopy class corresponding
to the representative path.
Thus we obtain
\begin{equation}
<L(\Delta x)> = \frac{ \sum_{C_{r}} L_{C_{r}} \; \exp[i S[n_{h}] /\hbar ]  }
{ \sum_{C_{r}} \; \exp[i S[n_{h}] /\hbar ]  },
\label{eq:operdeflength}
\end{equation}
where
\begin{equation}
\exp[ i S[n_{h}/\hbar ]
= K^{free}_{n_{h}}
\exp \left[ \frac{i q}{2 \pi \hbar c}
[ (\theta' -\theta) \phi_{tot}
+ 2\pi [ n_{1} \phi_{1} + \cdots n_{N_{S}} \phi_{N_{S}} ] \right],
\end{equation}
and $K^{free}_{n_{h}}$ has been determined from our experiment.
This yields $< L >$ for a given array of solenoids,
characterized by $\Delta x$, in the presence of the vector potential of the
solenoids. But we can as well obtain the length in absence of the vector
potential, i.e., corresponding to free propagation:
The length is stil given by Eq.(~\ref{eq:operdeflength})). But the action
weight factor, putting all fluxes $\phi_{i} \equiv 0$, is then
\begin{equation}
\exp[ i S[n_{h}/\hbar ]
= K^{free}_{n_{h}}.
\end{equation}
Finally, in order to extract the Hausdorff dimension $d_{H}$, we
have to measure the length
$<L(\Delta x)>$ for many values of $\Delta x $,
look for a power law behavior when $\Delta x \rightarrow 0$ and determine the
critical exponent and thus $d_{H}$.

\bigskip

At the end of this section let us discuss limitations and errors.
As a consequence of the fact that this experiment is not sensitive to the
zig-zagness parallel to the solenoids, we do not measure the length of the path
but only its projection onto the plane perpendicular to the solenoids, i.e., in
$D=2$ dimensions.
Nevertheless, the length as such is physically not so interesting
(it depends on $\Delta x$ anyway).
The physically important quantity is the critical exponent (Hausdorff
dimension)
which corresponds to taking the limit $\Delta x \rightarrow 0$.
But the latter should be the same in any space dimension.

\bigskip

Let us comment on the two length definitions employed here:
Firstly, there is the general definition, Eq.(~\ref{eq:gendeflength})
expressed as a path integral. Secondly, there is an operational definition
Eq.(~\ref{eq:operdeflength}) proposed to be applied in the experiment.
These definitions differ in two aspects:
(i) While Eq.(~\ref{eq:gendeflength}) describes the continuum limit and
takes into account infinitely many paths, Eq.(~\ref{eq:operdeflength})
is an approximation, taking into account only a finite number of paths.
This is the same kind of approximation as is done when replacing an ordinary
integral by a finite sum. When letting $\Delta x$ go to zero, the number of
paths tends to infinity. (ii) The more serious difference is the fact that
while Eq.(~\ref{eq:gendeflength}) represents all paths,
Eq.(~\ref{eq:operdeflength}) chooses only one representative path
for each homotopy class, although there may be several topologically
inequivalent paths in the same homotopy class. An example is shown in Fig.[15].

\bigskip

Finally, let us suppose the experiment were performed and
the experimentators would have extracted $d_{H}=1.5 \pm 0.1$. What would one
conclude from this? Firstly, such a result would establish that quantum paths
are fractals (anything with $d_{H} > 1$). But then one would ask: For
theoretical reasons, one would expect to find $d_{H}=2$. What is the reason for
this difference? As pointed out above, the theoretical result $d_{H}=2$
is not rigorously established. Thus it may be that $d_{H}=1.5$ is the correct
answer and $d_{H}=2$ is wrong. However, this is quite unlikely.
Nevertheless, it is very desirable to get a firm theoretical answer. This
could be done by proving Eq.(13) or at least by solving the path integral
by computer simulation in real time and extracting $d_{H}=2$.
Supposing, the latter had been achieved. Then one would conclude that the
proposed experiment is incomplete, in the following sense:
It shows that quantum paths are zig-zag, but the experiment "does not capture
all zig-zagness". Possible errors are: (a) The upper limit of winding number
imposed in the experiment is too small. (b) The distance between solenoids
$\Delta x$ is too large. (c) The chosen definition of representative path
in a given homotopy class, namely the shortest connected path among
topologically inequivalent paths within a homotopy class, is incorrect.
Errors (a) and (b) are questions of experimental precision, which may be a
substantial problem when doing a real experiment. However, we do not consider
it as a serious conceptual problem (in a Gedanken experiment).
The most serious point and possibly the weakest part in our view is (c).
If the definition of the representative path is incomplete, one can not test
quantum mechanics. In order to verify the proposed length definition
of a representative path, we suggest to do a numerical simulation of path
integrals in the presence of the solenoids and to compute the quantum
mechanical length as defined by Eq.(46) with that defined by Eq.(45).

\section{Summary}
We have made a proposal how to observe experimentally the zig-zag
motion of quantum mechanical paths. We have suggested two experiments:
Experiment I is conceived to show the existence of zig-zagness. Experiment II
is conceived to determine the average length of path, its scaling behavior when
$\Delta x \rightarrow 0$, and eventually the corresponding Hausdorff dimension.
Finally we want to return to the Heisenberg uncertainty principle, as discussed
in sect. 3 in the context of the multi-slit experiments. It says: The higher
the precision in measuring positions to determine a quantum path, the more will
it change the interference pattern of the quantum amplitude. What relevance
does this have to our experiment?
What does our experiment have in common with the multi-slit experiment of
sect.3? Where does it differ?
Firstly, in the multi-slit experiment the spatial resolution
is given by the distance between holes and the wave length of light. In our
experiment the spatial resolution is given by the distance between solenoids.
Thus the array of multiple screens with multiple holes corresponds to the array
of solenoids. This is a common feature between our experiment and the
multi-slit experiment. However, the rest is different: In the multi-slit
experiment,
by interaction with light, one determines the holes passed by the electron, but
at the same time modifies its quantum path and hence its interference pattern.
In our experiment, for each set of solenoids with a given set of fluxes,
corresponding to time-independent vector potentials, one does not measure the
position, but the undisturbed interference pattern of the electron
propagating in the vector potential. Then one changes the set of fluxes
(but not the position of solenoids), and measures again the interference
pattern. This is repeated a number of times. The crucial point is: Changing the
fluxes of the solenoids (and hence the vector potential) does not change the
free propagator in the homotopy class given by the array of solenoids.
Thus measuring the interference pattern for a large enough number of sets of
different fluxes allows to determine the free propagator in the homotopy
classes. From that follows the length. Thus contrary to the multi-slit
experiment, we do not determine the holes (or pairs of solenoids)
which the electron has passed, but relative amplitudes for classes of quantum
mechanical paths.

\bigskip

Finally, let us comment on practical feasibility of our proposed experiments.
Certainly experiment I is much simpler. Assuming the diameter of a realistic
small solenoid to be in the order of $d_{sol} = 5 \; \mu m$, we have seen that
quantum mechanical effects should be clearly observable when the solenoid is
placed in a wide range of distances (e.g., $h = d_{sol}$ to $h = 100 \;
d_{sol}$
for the particular choice of the parameters) from the classical trajectory,
provided that the flux is such that $\alpha$
takes values in the neighbourhood of half integer numbers.
The second experiment is more difficult. However, part of the experiment can
tested by comparison with theory: For the case of one single solenoid,
the homotopy classes and the free propagator in each homotopy class
are known analytically. This can be compared with the results of the
experiment.

\newpage
\begin{flushleft}
{\bf Acknowledgement}
\end{flushleft}
The author is grateful for discussions with Prof. Franson, Johns Hopkins
University, Prof. Pritchard, MIT, Prof. Cohen-Tannoudji, Ecole Normale
Sup\'erieure, and Prof. Hasselbach, Universit\"at T\"ubingen.
The author is grateful for support by NSERC Canada.

\newpage
\begin{flushleft}
{\bf Figure Caption}
\end{flushleft}
\begin{description}
\item[{Fig.1}]
Typical paths of a quantum mechanical particle
are highly irregular on a fine scale, as shown in the sketch.
Thus, although a mean velocity can be defined, no
mean square velocity exists at any point. In other words, the paths are
non-differentiable. Fig. taken from Ref.\cite{kn:Feyn65}.
By courtesy of McGraw Hill Publ.
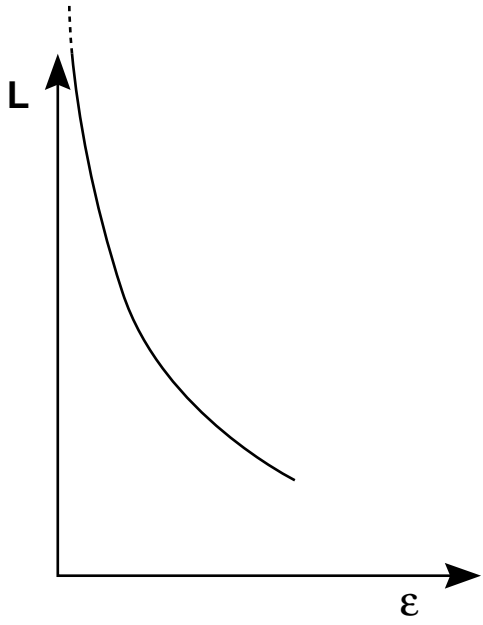
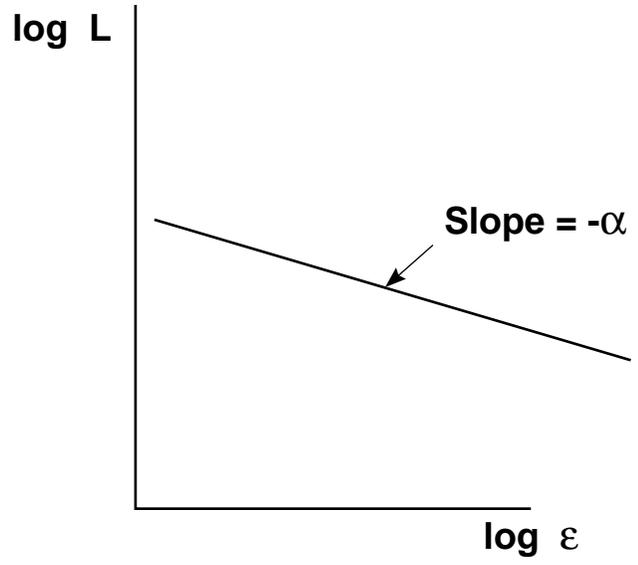
\item[{Fig.2}]
Graph corresponding to Eq.(17). Length versus $\epsilon$.
The log-log plot allows to determine the critical exponent $\alpha$.
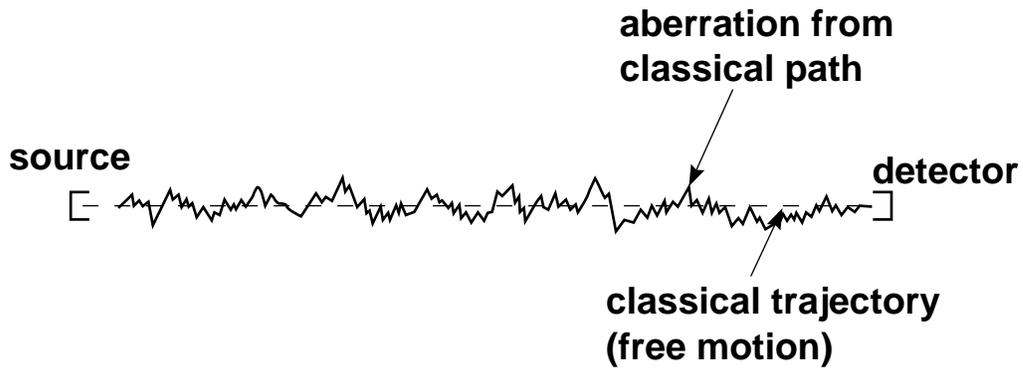
\item[{Fig.3}]
Schematic plot of propagation of massive free particle.
Dashed line: classical trajectory. Full line:
quantum mechanical zig-zag motion. Note: there is zig-zagness
in all spatial dimensions, i.e., in longitudinal as well as in
transversal directions.
\item[{Fig.4}]
Measurement of position and definition of elementary length scale $\Delta x$.
(a) Sequence of screens with holes. (b) Wire chamber.
\item[{Fig.5}]
Measurement of position of holes traversed by an electron.
(a) Determination of position by electron-light scattering.
(b) Determination of position by measuring recoil direction and momentum
of movable screen.
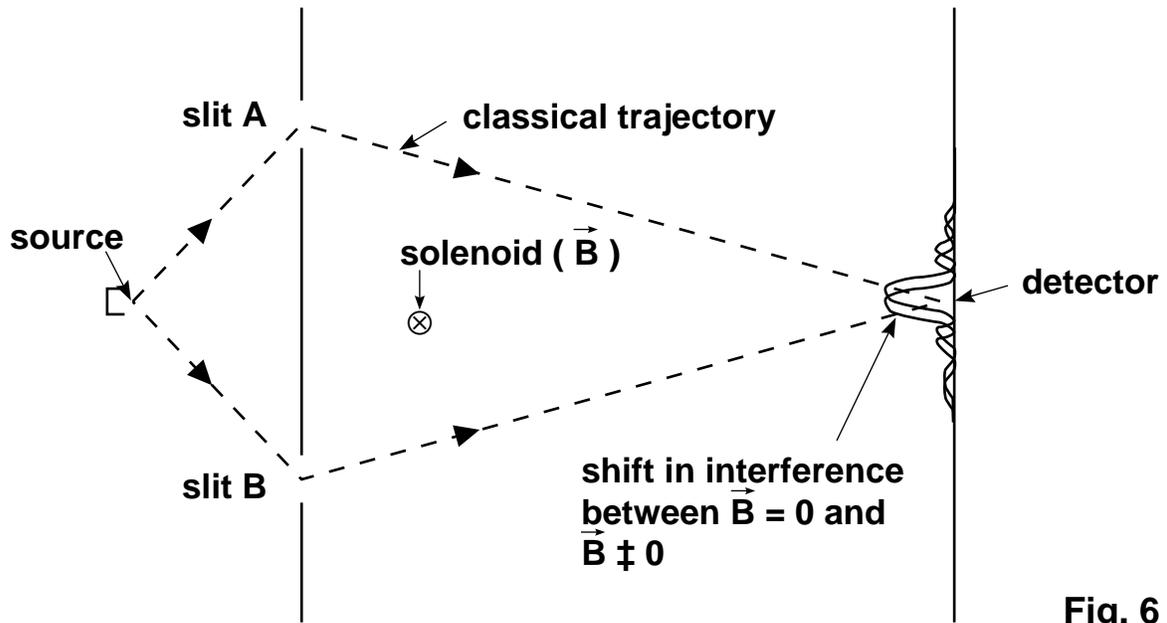
\item[{Fig.6}]
Set-up of Aharonov-Bohm experiment. A charged particle (electron)
is scattered from two slits and one observes interference
in the detector. When placing a solenoid (thin magnetic flux tube)
in the region between the classical trajectories, one observes a shift in
interference.
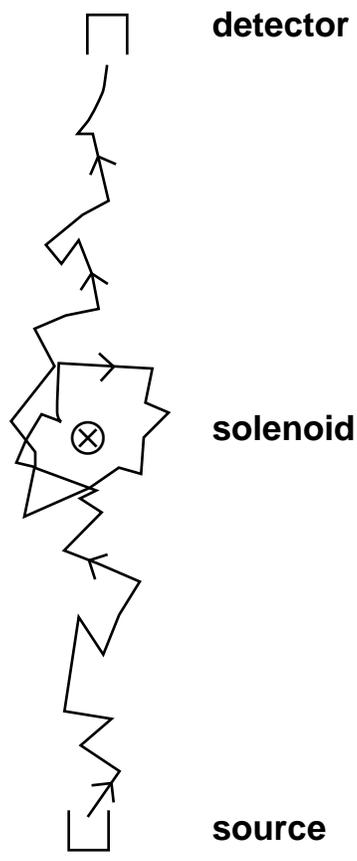
\item[{Fig.7}]
Schematic quantum path from source to detector winding around solenoid.
\item[{Fig.8}]
Free propagator. Comparison of exact result, Eq.(25), with
expansion in Bessel functions, Eq.(28). (a) Real part as function of distance
$h$ from classical path and of cut-off $m_{max}$, $5 \leq m_{max} \leq 15$.
(b) Same for imaginary part. Other parameters see text.
\item[{Fig.9}]
(a) Real part of semi-classical propagator, Eq.(32).
(b) Real part of Aharonov-Bohm propagator, Eq.(31). (c) Absolut value of
difference of both.
Dependence on  distance $h$ and on $\alpha$.
Other parameters as in Fig.[8], cut-off $m_{max}=50$.
\item[{Fig.10}]
Same as Fig.[9], but for imaginary part.
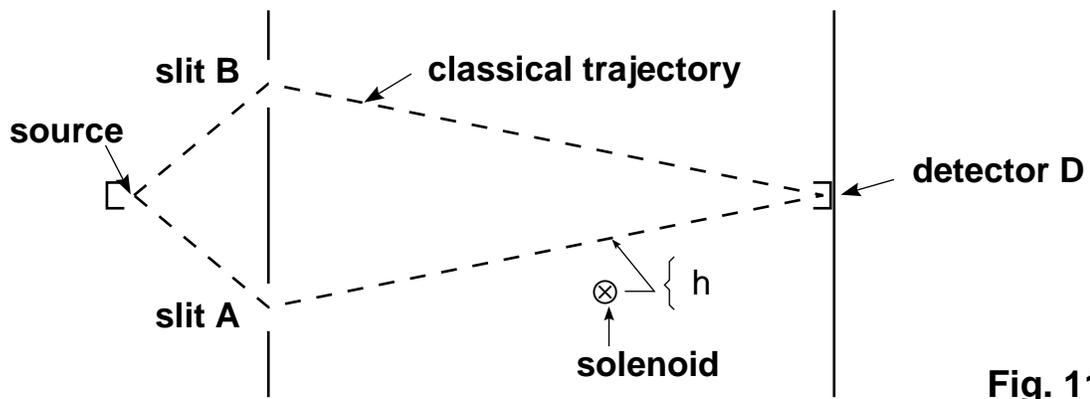
\item[{Fig.11}]
Set-up of Gedanken experiment I. Similar to Aharonov-Bohm experiment, but
the solenoid is placed outside the region bounded by the classical
trajectories.
\item[{Fig.12}]
Gedanken experiment I, similar to Fig.[11]. Sketch of interference pattern
if there were only two contributing quantum mechanical paths (in the
neighbourhood of the two classical paths). If the solenoid is situated outside
of the region bounded by the two quantum paths, there is no shift in
interference (a), otherwise there is a shift (b).
\item[{Fig.13}]
Like Fig.[9], after division by factor $\mu/2 \pi i \hbar T$.
Physical parameters see text.
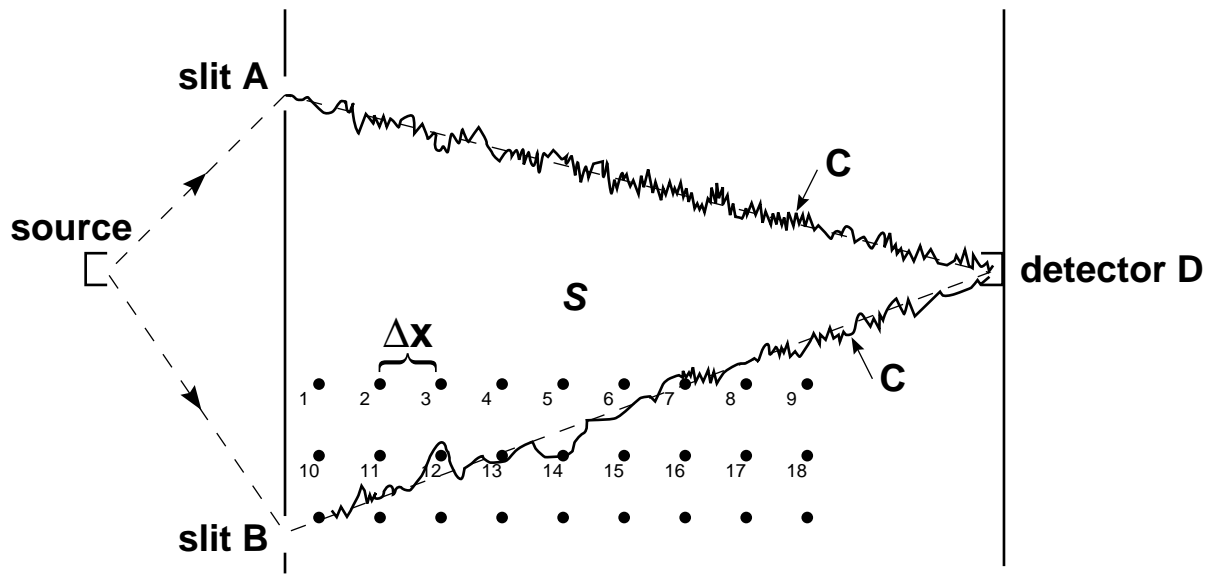
\item[{Fig.14}]
Set-up of Gedanken experiment II. Like Gedanken experiment I, but there are
many solenoids.
\item[{Fig.15}]
Topologically equivalent paths.
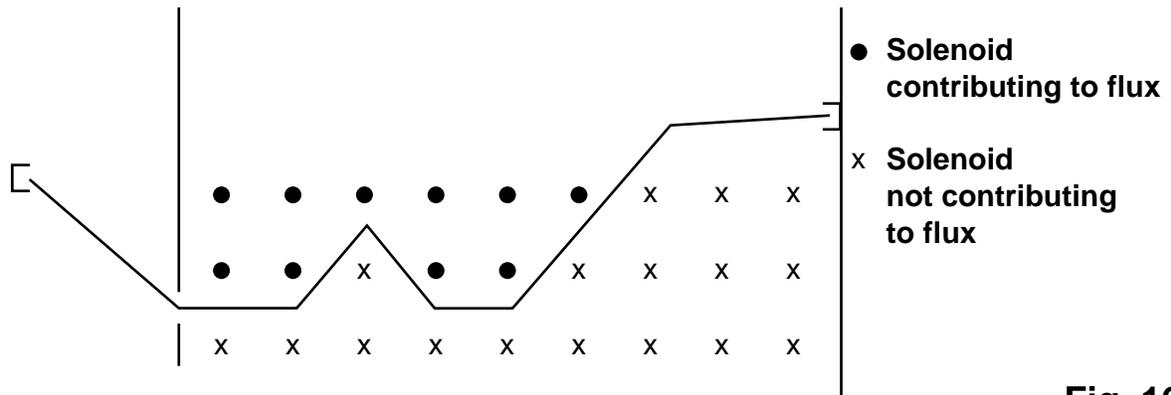
\item[{Fig.16}]
Gedanken experiment II. Prescription how to associate a length to a particular
quantum path corresponding to Fig.[14].
\end{description}


\newpage
\begin{thebibliography}{999}
\bibitem{kn:Feyn65} R.P. Feynman and A.R. Hibbs,
Quantum Mechanics and Path Integrals, McGraw Hill, New York (1965).
\bibitem{kn:Mand83} B.B. Mandelbrot, The Fractal Geometry of Nature, Freeman,
New York (1983).
\bibitem{kn:Abbo81} L.F. Abbot and M.B. Wise,
Am. J. Phys. 49(1981)37.
\bibitem{kn:Camp82} E. Campesino-Romeo, J.C. D'Olivio and M. Sokolovsky, Phys.
Lett. A89(1982)321.
\bibitem{kn:Feyn65a} R.P. Feynman and A.R. Hibbs,
Quantum Mechanics and Path Integrals, McGraw Hill, New York (1965), sect. 7.2
and 7.3.
\bibitem{kn:Krog95} H. Kr\"oger, S. Lantagne, K.J.M. Moriarty, and B. Plache,
Phys. Lett. A199(1995)299.
\bibitem{kn:Made81} O. Madelung, Solid State Theory, Springer,
Berlin (1981).
\bibitem{kn:Brue55} K.A. Brueckner, Phys. Rev. 97(1955)1353.
\bibitem{kn:Ahar59} Y. Aharonov and D. Bohm, Phys. Rev. 115(1959)485.
\bibitem{kn:Wilc90} F. Wilczek, Fractional Statistics and Anyon
Superconductivity, World Scientific, Singapore (1990).
\bibitem{kn:Wern60} F.G. Werner, D.R. Brill, Phys. Rev. Lett. 4(1960)344;
see Refs. therein.
\bibitem{kn:Feyn64} R.P. Feynman, R.B. Leighton and M. Sands,
The Feynman Lectures on Physics, Addison Wesley, Reading (1964), Vol.II.
\bibitem{kn:Schu71} L.S. Schulman, J. Math. Phys. 12(1971)304.
\bibitem{kn:Schu81} L.S. Schulman, Techniques and Application of Path
Integration, John Wiley, New York (1981).
\bibitem{kn:Tono83} A. Tonomura, H. Umezaki, T. Matsuda, N. Osakabe, J. Endo,
Y. Sugita, Phys. Rev. Lett. 51(1983)331; see Refs. therein.
\bibitem{kn:Chen84} T.P. Chen, L.F. Li, Gauge Theory of Elementary Particle
Physics, Oxford University Press, Oxford (1984).

\end{thebibliography}
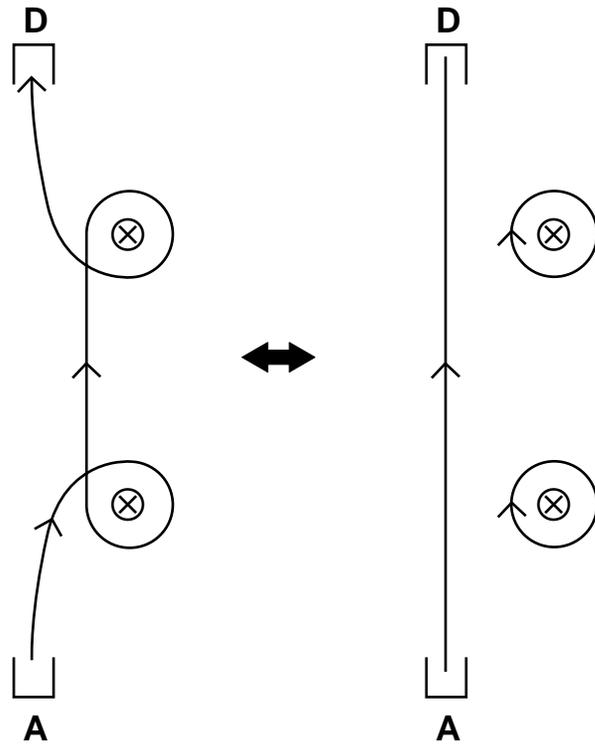
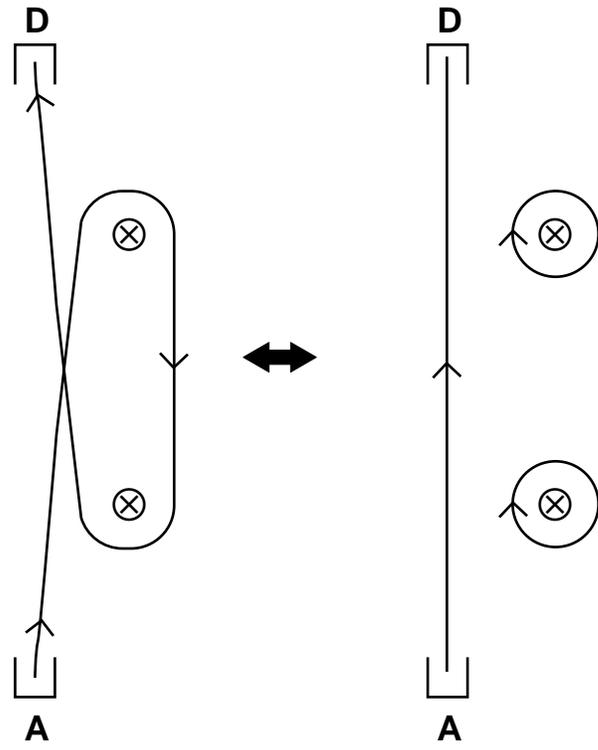
\end{document}